\def\BEDOS{BEDOS}

\def\IAUPrinc{IAU Symp.\ No.\ 127, {\it Structure and Dynamics of Elliptical
   Galaxies\/} (Princeton), ed.\ T.\,de~Zeeuw, Reidel: Dordrecht}

\def\LaSerena{{\it Bulges of galaxies\/} (La Serena), eds. B.\ Jarvis and
D.M.\,Terndrup, ESO Workshop Publ.: Garching}
\def\Austin{{\it Photometry, Kinematics, and Dynamics of Galaxies\/}, (Austin),
ed. D.S.\,Evans, Texas Univ.\ Press: Austin}

\def\MC{Ca\-pac\-cio\-li}
\def\eppoi{\&}
\def\CHN{\MC\ \etal}
\def\ApJ{{\it ApJ\/}}
\def\ApJL{{\it ApJL\/}}

\def\AJ{{\it AJ\/}}
\def\AsAp{{\it A{\rm \&\/}A\/}}

\def\MN{{\it MNRAS\/}}

\def\Messenger{{\it The Messenger\/}}

\def\BGD{\begin{description}}
\def\EDD{\end{description}}
\def\BGF{\begin{figure}}
\def\EDF{\end{figure}}
\def\BGC{\begin{center}}
\def\EDC{\end{center}}
\def\BGT{\begin{tabular}}
\def\EDT{\end{tabular}}
\def\BGE{\begin{equation}}
\def\EDE{\end{equation}}

\def\rms{\mbox{\it r.m.s.\/}}
\def\kms{\mbox{km~s$^{-1}$}}

\def\ML{\mbox{$M/L$}}
\def\secondip{\hbox{\rlap{\hbox{.}}\hbox{$''$}}}
\def\primip{\hbox{\rlap{\hbox{.}}\hbox{$'$}}}
\def\gradip{\hbox{\rlap{\hbox{.}}\raise 5.truept \hbox{{\small$\circ$}}}}
\def\mincir{\ \raise-2.truept\hbox{\rlap{\hbox{$\sim$}}\raise5.truept
\hbox{$<$}\ }}
\def\magcir{\ \raise-2.truept\hbox{\rlap{\hbox{$\sim$}}\raise5.truept
\hbox{$>$}\ }}
\def\etal{{\it et al.\/}}
\def\cf{{\it cf.\/}}
\def\ie{{\it i.e.\/}}
\def\eg{{\it e.g.\/}}

\def\vass{\mbox{$\langle v_f\rangle$}}

\documentstyle{l-aa}        
\begin{document}

\thesaurus{03(11.11.1 NGC~3115)}
\title{Deep kinematics and dynamics of edge--on S0 galaxies. I.
NGC~3115\thanks{Based on observations carried out at ESO, La Silla, Chile}}
\author{M.\ Capaccioli\inst{1}\inst{2} \and E.\ Cappellaro\inst{3} \and
E.V.\ Held\inst{4} \and M.\ Vietri\inst{5}}
\institute{Dipartimento di Astronomia, Universit\`a di Padova,
Vicolo dell'Osservatorio 5, I--35122 Padova, Italy \and
Osservatorio Astronomico di Capodimonte,
Vicolo Moiariello 16a, 80131 Napoli, Italy \and
Osservatorio Astronomico di Padova
Vicolo dell'Osservatorio 5, I--35122 Padova, Italy \and
Osservatorio Astronomico di Bologna,
Via Zamboni 33, I--50125 Bologna, Italy \and
Osservatorio Astronomico di Roma,
Via dell'Osservatorio, I--00040 Monte Porzio Catone (Roma), Italy}

\date{Received ; accepted }

\maketitle
\begin{abstract}
As a first step of a program aimed to the detection of dark matter (or radial
variations of \ML) in early--type galaxies, we report  deep spectroscopic
observations of the bulge--dominated edge--on S0 galaxy NGC~3115, made at ESO,
La Silla, using EFOSC at the 3.6\,m telescope and EMMI at NTT.  Such
observations allow measurements of the rotational velocity out to $1.8 a_e$
(effective radii) from the galaxy center, where the surface brightness is
$\mu_B \simeq 24$ mag arcsec$^{-2}$.  The rotation curve quickly reaches an
asymptotic value, $\vass \simeq 260$ \kms,  with only marginal indication of
systematic decline within the range of our observations.  The line--of--sight
velocity dispersion has also been measured; it decreases steeply from a rather
high central value  and flattens  out  ($\langle\sigma\rangle \simeq 100$
\kms)  within  our observing range ($a \mincir 1.3 a_e$).

Models built on these data and simple dynamical arguments show that the \ML\
of NGC~3115 must thus be increasing from $\ML = 6$ (in solar units) in the
inner regions ($\sim 1 a_e$) to at least $\ML \geq 10$ in the outermost regions
($\sim 2 a_e$).

\end{abstract}

\section{Introduction}
Some evidence for the existence of dark matter has now been acquired on all
scales, from those of dwarf galaxies up to the richest galaxy clusters.
At the same time inflationary theories suggest that dark matter might exist
even at cosmological scales.
However, at galactic scales, that is at galactocentric distances comparable
with the lengths of the effective radii $a_e$, such evidence is still patchy
(\cf\ Kormendy \eppoi\ Knapp 1987).
It seems rather firm for dwarf and spiral galaxies, whereas no direct evidence
has yet been discovered for early--types (E and S0's).
This happens primarily for three reasons.
In the first place, early--type galaxies have little, if any interstellar
matter (Knapp 1987), which traces the rotation curve and, indirectly, the dark
matter in spirals;
the ionized gas in early--type galaxies is mostly confined to the innermost
regions, where dark matter is unlikely to be dominant, and all this gas has a
very uncertain geometry, which severely limits the interpretation of
observations.
Cooling flows in ellipticals provide some support for dark matter at very
large radii (Fabian, Arnaud \eppoi\ Thomas 1987), but these estimates are
fraught with uncertainties (Bertin, Pignatelli \eppoi\ Saglia 1992).
Furthermore they apply only to a small and somewhat peculiar subset of all
ellipticals, the brightest cluster members, which all happen to be ``boxy''
galaxies (Bender \etal\ 1989, \MC, Caon \eppoi\ D'Onofrio 1992).
For the other class of ellipticals, the ``disky'' E's, which constitute
$\sim50\%$ of all E's (Bender \etal\ 1989), nothing is known.

Bulge--dominated edge--on S0's (hereafter referred to as \BEDOS) provide a
convenient laboratory to investigate the presence of dark matter inside the
optical regions of early--type galaxies.
First of all, bulge dominated S0's are now believed to be closely related to
the class of ``disky'' ellipticals (\MC\ 1987, 1990, \MC, Caon \eppoi\ Rampazzo
1990) for which even the equivocal evidence provided by the X--ray coronas is
not available.
Moreover, the existence of dark matter haloes surrounding S0 galaxies seems to
be demanded by those few such objects possessing polar rings (\eg\ Sparke 1989
and Arnaboldi \etal\ 1992). In this respect, the study of \BEDOS\ by direct
kinematical mapping along the major axis is greatly simplified for the
following reasons:

1.\ \ they have a high $v_{max}/\sigma_c$ ratio (\MC\ 1979, Davies \etal\
1983), which makes the analysis simpler (and more similar to that of spirals);

2.\ \ the geometry is known since, as $i \simeq 90^\circ$, the apparent axial
ratio equals the intrinsic one;

3.\ \ the bulge is unlikely to be strongly triaxial (for NGC~3115 and NGC~3379,
see \MC\ \etal\ 1991);

4.\ \ direct observations provide the photometric characteristics and the \ML\
of the disk (and thus its dynamical relevance), which allow us to subtract its
effects;

5.\ \ thanks to the ``cylindrical rotational field'' (i.e. very slow variation
of the velocity perpendicularly to the galaxy major axis; \cf\ Kormendy
\eppoi\ Illingworth 1982), we know how to obtain the true rotation velocity of
the bulge, without contamination from the disk;

6.\ \ lastly, substantial photometric and spectroscopic literature exists,
which permits relevant checks and complements modern observations.
\medskip

\noindent
A kinematical project involving \BEDOS\ may be simpler than considering
directly elliptical galaxies, but it is not altogether simple.
In fact, detection of dark matter requires spectroscopic observations at
large distances from the centers of galaxies.
As a way of comparison, existing models of the mass distribution of the Galaxy
(Caldwell \eppoi\ Ostriker 1981, and Bahcall, Schmidt \eppoi\ Soneira 1982)
show that, within the solar radius (8.5~kpc, corresponding to $\sim2.5$ disk
scale--lengths), the mass is roughly equally distributed between spheroid,
disk, and dark halo.
For a similar distribution of matter in early--type galaxies, one should find
$\simeq1/2$ of the total mass in dark matter within $\simeq 1.8$ effective
radii $a_e$, the radial range corresponding to the same fraction of the total
light in the Milky Way.
Spectroscopic observations of stellar lines at these distances from the center
of the galaxy are hampered by the faintness of the continuum, which is
notoriously already faint enough at $1\;
a_e$:
$\mu_e \simeq 22.5\pm2$ B-mag arcsec$^{-2}$ for E galaxies and S0 bulges with
total luminosity $M_B \mincir -18$ (\MC\ \eppoi\ Caon 1991; their Fig.~1).

Kinematical data out to galactocentric distances of $\sim 2 a_e$ have now been
obtained for the first time as part of a program started at ESO, La Silla, in
1989, and aimed at investigating the dependence of the dynamical behavior of
galactic disks on the bulge--to--disk ratios, and the presence and shape of
dark haloes in early--type galaxies (Cappellaro, \MC\ \eppoi\ Held 1989, 1990)
through observations of  ``deep'' rotation and velocity dispersion curves in
\BEDOS\ and polar ring bulges (Arnaboldi \etal\ 1992).
This program has been made possible by the unique capabilities of
the ESO Faint Object Spectrograph and Camera (EFOSC; Melnick, Dekker \eppoi\
D'Odorico 1989) attached to the Cassegrain focus of the ESO 3.6\,m telescope.
The first target galaxies, NGC~2310, NGC~3115, and NGC~4179, were chosen among
the few \BEDOS\ with fairly large angular sizes (distances $\le25$~Mpc) with
the purpose of minimizing resolution problems.

Here we concentrate on NGC~3115, leaving the remaining objects for a following
paper. One reason for this choice is that much work on this ``standard'' S0
galaxy has already been carried out (see \MC, Held \eppoi\ Nieto 1987, for an
extensive photometric mapping, and $\S$2 for references on kinematics and
dynamics). In particular, we already know the mass--to--light ratio of the
exponential disk of this galaxy (\MC, Vietri \eppoi\ Held 1988) and we have
also shown that the dominating bulge is unlikely to be very triaxial (\MC\
\etal\ 1991).
Another reason is our need to demonstrate, with a good example, the
reliability of the EFOSC data, particularly for what concerns the velocity
dispersion measurements (see $\S$2), taking advantage of the higher resolution
but more expensive results from spectra taken with EMMI at the ESO NTT.

The plan of the paper is as follows.
Section~2 details the spectroscopic observations and their reduction, with
results doubling the radial extension of the published kinematical measurements
in this very well studied galaxy.
Major--axis rotation and velocity dispersion curves of NGC~3115 are analyzed
theoretically in Sect.~3.
Section~4 contains a summary and our conclusions.
An Appendix discusses the theoretical method employed for the dynamical
analysis, and expands on a few technical details.
The relevant data on NGC~3115, mostly taken from \CHN\ (1987), are summarized
in Table~1.

\begin{table*}
\begin{tabular}{lll}
\multicolumn{3}{c}{{\large\bf Table 1}}\\[12pt]
\multicolumn{3}{c}{NGC~3115: catalog data}\\[7pt]
\hline\hline\\[-7pt]
Parameter & Symbol & Value \\[5pt]\hline\\[-7pt]
Right Ascension \ \,(1950.0) & $\alpha$ & $10^h\ 02^m\ 44^s.51$ \\
Declination \ \ \ \ \ \ \ \
(1950.0) & $\delta$ & $-07^\circ\ 28'\ 30\secondip7$\\
Morphological type && S0$^-$(sp)\\
Standard isophotal major diameter & $D_{25}$ & $8\primip63 \pm 0\primip27$ \\
Standard axis ratio & $(b/a)_{25}$ & $0.51 \pm 0.03$ \\
Standard position angle & P.A.$_{25}$ & $43\gradip5 \pm 1^\circ$ \\
Corrected total magnitude$^\ast$ & $B_T^{\circ}$ & $9.65 \pm 0.1$ \\
Observed radial velocity $[\kms]$\dag & $v_0$ & 663\\
Adopted distance [Mpc] & $\Delta$ & $10$ \\
Scale factor [pc] & $1''$ & $48.5$ \\
Central rotational vel. gradient [\kms\ arcsec$^{-1}$]\dag &
$\Big(dv(a)/da\Big)_{a=0}$ & $>100$\\
Asymptotic rotational velocity [\kms]\dag & $\vass $ & 260 \\
Central velocity dispersion [\kms]\dag & $\sigma(0)$ & 325 \\
Velocity dispersion at $1a_e$ [\kms]\dag & $\sigma(a_e)$ & 105 \\
{\it Spheroid:\/} \\
\hspace{1cm}Effective semi--major axis & ${a_e}$ & $93'' \pm 8''$ \\
\hspace{1cm}Effective semi--minor axis & ${b_e}$ & $35'' \pm 3''$ \\
\hspace{1cm}Corrected total magnitude & $(B_T^{\circ})_b$ & $9.71 \pm 0.1$ \\
\hspace{1cm}Mass--to--light ratio at $1a_e$ [solar=1]\dag & $\ML$ & 6 \\
\hspace{1cm}Mass--to--light ratio at $2a_e$ [solar=1]\dag & $\ML$ & $\ge10$ \\
{\it Disk:\/} \\
\hspace{1cm}Fractional luminosity & $k_D$ & $0.06 \pm 0.01$ \\
\hspace{1cm}Face--on scale length$\ddag$ & $h$ & $25\secondip5$ \\
\hspace{1cm}Inclination$\ddag$ & $i$ & $86^\circ$ \\
\hspace{1cm}Mass--to--light ratio [solar=1]$\ddag$ & $\ML$ & 7 \\
[2pt]\hline\\[5pt]
\multicolumn{3}{l}{$^\ast$ From $B_T=9.75$ (\CHN\ 1987) with $A_B=0.1$
(Burstein \eppoi\ Heiles 1982).}\\
\multicolumn{3}{l}{$\dag$ This paper.}\\
\multicolumn{3}{l}{$\ddag$ From \MC, Vietri \eppoi\ Held (1988).}
\end{tabular}
\end{table*}

\section{Spectroscopic observations}
\subsection{The EFOSC spectra}
Five long exposure spectra of NGC~3115 (Table~2) were obtained with EFOSC at
the ESO 3.6\,m telescope, and
recorded with a high resolution RCA CCD ($640\times1024$ pixels; ESO
code \#8), in binned mode: pixel size~$=30\mu$m, or $0\secondip675$ on
the sky.
We chose the ESO {\it Orange 150\/} grism, with dispersion of
$3.9\,$\AA~px$^{-1}$ from $5000$ to $7000\,$\AA.
The $3\primip6\times1\secondip5$ spectrograph slit was aligned with the major
axis of NGC~3115 (P.A.$= 43\gradip5$).
Although the nucleus of the galaxy was set close to one end of the slit, the
other end still viewed a region of the object as bright as $\mu_B
\simeq24.5$, \ie\ only $\sim7$~times fainter than the average night--sky
brightness.
Therefore we also obtained long exposure spectra of the blank night--sky
(by offsetting the telescope by $\sim20'$), in chronological sequence with the
astrophysical exposures.

The raw data were pre--processed by the standard procedures in
MIDAS (ESO Operating Manual No.~1) for bias and dark
subtraction, and for flat--fielding based on dome and dawn--sky
exposures at different count levels.
The complex distortion pattern was mapped using the comparison spectra to
derive a line--by--line wavelength calibration.
We found that the position of the comparison lines drifts over the detector
during the night, the amplitude of the displacement being of the order of
1~pixel (equivalent to $\sim200$ \kms).
The consequent frame--to--frame shift was corrected by matching the position
of the night--sky emissions.
The drift acted on the zero point of the absolute velocity scale, adding an
uncertainty of $\pm50$ \kms\ (\rms) to the systemic velocity.
More importantly, it broadened significantly the long exposure spectra of the
galaxy, thus introducing a difference with the very short exposure spectra of
the template stars (which did not suffer any significant widening by the
drift).
Repercussions of such a difference on velocity dispersion measurements will be
considered below (Sect.~\ref{SSvel_disp}).

A crucial step in the reduction was the night--sky subtraction
from galaxy spectra.
We used the blank sky exposures since no part of
the slit in our spectra of NGC~3115 was free from the galaxy signal.
The frames corresponding to the same side of the galaxy major axis were
individually sky subtracted, and then averaged with an algorithm
correcting for cosmic ray events.

The kinematical data were analyzed with a package (Bender 1990)
based on the cross--correlation (CC), at each radial bin, of the
continuum--subtracted galaxy spectrum with a template.
No differences were found by changing the template stars
(spectral classes from G8 to K1).

\subsection{{\rm B\&\/C\/} and {\rm EMMI} spectra}
Given the large uncertainty in the zero point of the velocity
scale, and the off--centering of the galaxy nucleus at one end of each
spectrum (which, with the steep central gradient of the rotational
velocity, prevented the sliding fit of
the velocity curve and the search for the
systemic velocity by the folding technique ---~see below), and most
importantly, the extra--broadening of the galaxy spectra with respect to the
templates, we complemented the deep EFOSC material with other spectra
of NGC~3115 at higher resolution.

\begin{table}
\begin{tabular}{lccl}
\multicolumn{4}{c}{{\large\bf Table 2}}\\[12pt]
\multicolumn{4}{c}{NGC~3115: journal of observations}\\[7pt]
\hline\hline\\[-7pt]
Instrument            & Nucleus   & Exp.\ time    & \phantom{ooo}Date\\
                     & placed at & [seconds]     & \\[5pt]
\hline\\[-7pt]
ESO 3.6\,m $+$ EFOSC & SW        & 3600          & 1989 Feb. 7\\
ESO 3.6\,m $+$ EFOSC & SW        & 1536          & 1989 Feb. 7\\
ESO 3.6\,m $+$ EFOSC & SW        & 1800          & 1989 Feb. 7\\
ESO 3.6\,m $+$ EFOSC & NE        & 3600          & 1989 Feb. 8\\
ESO 3.6\,m $+$ EFOSC & NE        & 3000          & 1989 Feb. 9\\
ESO 3.6\,m $+$ B\&C  & center    & 1200          & 1990 Mar. 21\\
ESO NTT    $+$ EMMI  & NE        & 3600          & 1991 Dec. 30\\
ESO NTT    $+$ EMMI  & NE        & 3$\times$3600 & 1991 Dec. 31\\
[3pt]\hline\\[5pt]
\multicolumn{4}{l}{In all cases the slit has been oriented at ${\rm P.A.}
= 43\gradip5$}
\end{tabular}
\end{table}

In March 1990 we took a relatively `light' spectrum of
NGC~3115 with the Boller \eppoi\ Chivens (B\&C) spectrograph attached to the
Cassegrain focus of the ESO 3.6\,m telescope (Table~2), using grating \#10
which, in the second order, gives a dispersion of $59\,$\AA~mm$^{-1}$ over the
range 4755--5640$\,$\AA.
The $6'$ long slit was opened to $222~\mu$m ( $\equiv 1\secondip6$),
aligned with major axis and centered on NGC~3115 nucleus.
The spectrum of the galaxy (and of a few template stars) was recorded on the
same CCD~\#8 as above, spatially binned to $30~\mu$m., with
resolution FWHM$ \simeq 1.75$~\AA\ ($=2$ pix).
This material was preprocessed and reduced by the same technique applied to
the EFOSC spectra.
The systemic velocity, obtained by matching the two sides of the rotation
curve opposite to the center, is $v_0 = 663 \pm 5$ \kms, in agreement with
Rubin, Peterson \eppoi\ Ford (1980).
We note explicitly that the coordinate of the velocity center
coincides with that of the peak of $\sigma$, in agreement with the symmetry of
the velocity dispersion.

The B\&C data were not extended enough to solve all problems of the EFOSC
spectra. Thus four spectra of one hour each (Table~2) were taken in
December 1991 with
EMMI, a spectrograph attached to the Nasmyth focus of the ESO 3.5\,m New
Technology Telescope (NTT).
We used grating \#6 which gives a dispersion of $28\,$\AA~mm$^{-1}$ over the
range 4900--5400$\,$\AA.
The $6'$ long slit was opened to $1\secondip0$, aligned with major axis of
NGC~3115.
The galaxy nucleus was always set at $\sim100''$ from the NE end of the slit.
The spectra of the galaxy (and of two template stars) were recorded on ESO
CCD~\#18, with pixels of 19 $\mu$m, with
resolution FWHM$\simeq 1.4\,$\AA\ ($= 2.6$ pix).

The EMMI spectra were again analyzed like
the EFOSC spectra, and then co--added.
One important difference here is that, due to the relatively longer slit, we
performed the blank sky subtraction using the signal at the SW edge of
the co--added spectrum.
The systemic velocity from the co--added spectrum is $v_\circ= 647 \pm 4$
\kms\ (error not including zero point uncertainty).

\subsection{The rotation curve}
Radial velocity measurements for the combined EMMI spectra are plotted in
Fig.~1, after folding about the galaxy center and the systemic velocity.
The data were also averaged over intervals with amplitude increasing with the
galactocentric distance up to a maximum of $\Delta a = 10''$; the \rms\
deviation about the mean value never exceeds 10 \kms, even
in the most distant bin.

\begin{figure}
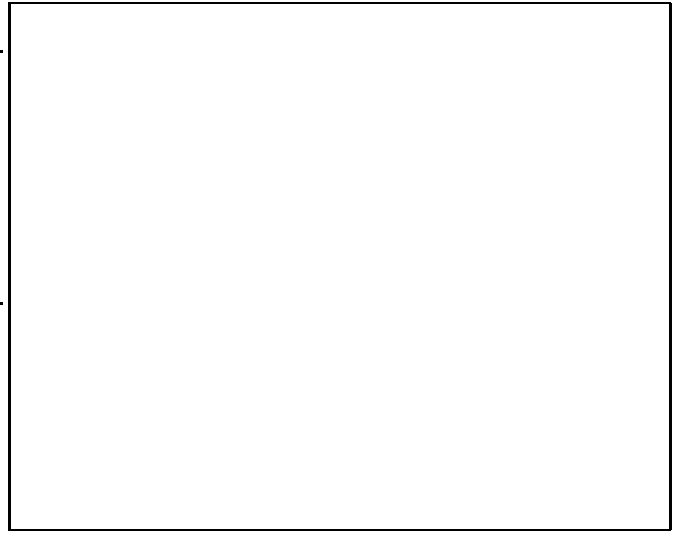

\picplace{7cm}
\caption{
Radial velocities $v(a)$ measured along the major axis of NGC~3115
and relative to the systemic velocity, plotted against the distance from the
center of the galaxy (counted positive on both sides opposite to the center).
The sign of the velocities along the SW semiaxis (where $v(a)<0$) has been
changed.
Coding for symbols is: EFOSC~= filled circles, EMMI~= filled diamonds.
The solid line, reproducing the plain interpolation of the two data sets,
runs quite flat from $a = 50'' \simeq a_e/2$ to $\sim 2a_e$. Representative
error bars are also shown.}
\end{figure}

The off--centered EFOSC spectra were combined using the B\&C and EMMI radial
velocity curves as a reference, the arbitrariness of the slide fitting
technique being minimized by the characteristic run of $v(a)$: a steep central
rise followed by a constant plateau.
The resulting mean velocities for the EFOSC spectra, plotted in Fig.~1, were
binned over $3''$ radial intervals from $a=19\secondip6$ and out to
$70\secondip5$, and over $10''$ intervals out to the record distance of
$\sim 2 a_e$.
The \rms\ deviations are larger than the internal error
estimates from the CC method (both for the radial velocities and for
the velocity dispersion measurements).
As an upper limit to the uncertainty of the individual measurements we
computed the standard deviations of the differences \mbox{$\Delta v(a)=
v_{NE}(a) - v_{SW}(a)$} and \mbox{$\Delta\sigma(a) = \sigma_{NE}(a) -
\sigma_{SW}(a)$}, obtaining, in both cases, a value of $\simeq 20$ \kms.

\begin{figure}
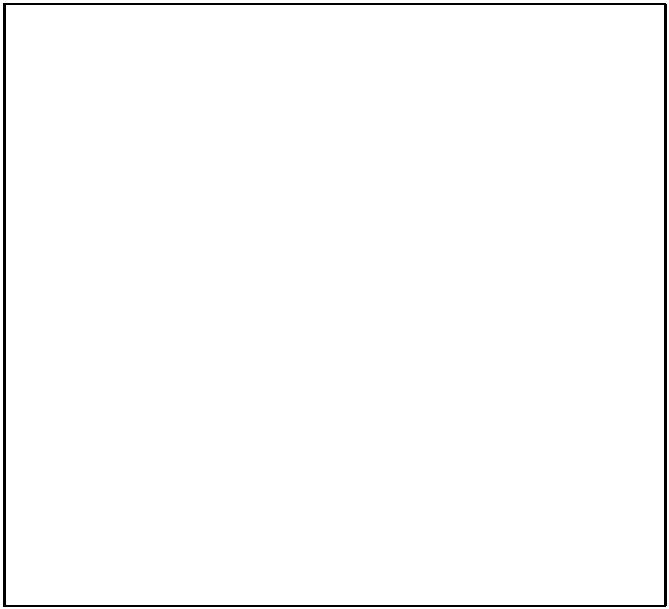

\picplace{8cm}
\caption{
Comparison of our radial velocity measurements (EFOSC~$=\times$, EMMI~$=+$,
B\&C~$=\circ$) with literature data: Illingworth \eppoi\ Schechter (1982)~=
filled circles, Rubin \etal\ (1980)~= filled squares (NE side) and
diamonds (SW side).
The steeper inner gradient shown by our data finds confirmation in the high
resolution rotation curve plotted in Kormendy's (1987); see also Fig.~6.}
\end{figure}

Visual inspection (Fig.~1) shows that the mean curve produced by the EFOSC
spectra ---~which has about the same resolution of the B\&C and EMMI curves in
the central region (Fig.~2)~--- runs essentially flat from $20''$ out to the
last observed point, with $\langle v\rangle = 262 \pm 9$ \kms\ (the quoted
uncertainty is the \rms\ dispersion).
Even if the EMMI data are fully consistent, within the errors, with the EFOSC
data, they suggest that $v(a)$ decreases slowly with the
galactocentric distance as: $v(a)=(252-0.09\times a)$ \kms, for $a\magcir
50''$.
The plain interpolation of the EFOSC and EMMI data, listed in Table~3 and
drawn in Fig.~1 as a solid line, runs flat at $\vass = 253\pm9$ \kms\ from $a
\magcir a_e/2$ out to $\sim 2 a_e$.

\begin{table}
\begin{tabular}{rrrcrrr}
\multicolumn{7}{c}{{\large\bf Table 3}}\\[12pt]
\multicolumn{7}{c}{NGC~3115: kinematical data [\kms]}\\[7pt]
\hline\hline\\[-7pt]
$a''$ & $v(a)$ & $\sigma(a)$ &\phantom{MM}&
$a''$ & $v(a)$ & $\sigma(a)$
\\[5pt]\hline\\[-7pt]
$  0 $&$   0 $&$ 302 $&&$  34 $&$ 260 $&$ 118 $\\
$  1 $&$ 103 $&$ 260 $&&$  37 $&$ 256 $&$ 113 $\\
$  2 $&$ 135 $&$ 232 $&&$  40 $&$ 250 $&$ 122 $\\
$  3 $&$ 157 $&$ 220 $&&$  43 $&$ 255 $&$ 107 $\\
$  4 $&$ 162 $&$ 212 $&&$  46 $&$ 250 $&$ 110 $\\
$  5 $&$ 172 $&$ 209 $&&$  49 $&$ 243 $&$ 109 $\\
$  6 $&$ 174 $&$ 204 $&&$  52 $&$ 253 $&$ 118 $\\
$  7 $&$ 177 $&$ 198 $&&$  55 $&$ 256 $&$ 122 $\\
$  8 $&$ 186 $&$ 190 $&&$  58 $&$ 257 $&$ 104 $\\
$  9 $&$ 205 $&$ 172 $&&$  61 $&$ 249 $&$ 107 $\\
$ 10 $&$ 211 $&$ 165 $&&$  64 $&$ 250 $&$ 107 $\\
$ 11 $&$ 217 $&$ 170 $&&$  67 $&$ 252 $&$ 104 $\\
$ 12 $&$ 224 $&$ 167 $&&$  70 $&$ 266 $&$  97 $\\
$ 13 $&$ 236 $&$ 160 $&&$  80 $&$ 251 $&$ 101 $\\
$ 14 $&$ 244 $&$ 153 $&&$  90 $&$ 252 $&$ 104 $\\
$ 15 $&$ 247 $&$ 153 $&&$ 100 $&$ 238 $&$ 104 $\\
$ 16 $&$ 255 $&$ 144 $&&$ 110 $&$ 261 $&$ 116 $\\
$ 17 $&$ 249 $&$ 147 $&&$ 120 $&$ 250 $&$ 110 $\\
$ 18 $&$ 252 $&$ 151 $&&$ 130 $&$ 238 $&\\
$ 19 $&$ 254 $&$ 144 $&&$ 140 $&$ 249 $&\\
$ 20 $&$ 269 $&$ 132 $&&$ 150 $&$ 250 $&\\
$ 22 $&$ 266 $&$ 129 $&&$ 160 $&$ 257 $&\\
$ 25 $&$ 265 $&$ 124 $&&$ 170 $&$ 274 $&\\
$ 28 $&$ 262 $&$ 119 $&&$ 180 $&$ 257 $&\\
$ 31 $&$ 256 $&$ 123 $&&$ 190 $&$ 250 $&\\
[3pt]
\hline\\[-7pt]
\end{tabular}
\end{table}

A comparison of our results with the other modern data available in tabular
form (Rubin \etal\ 1980, Illingworth \eppoi\ Schechter 1982) is also given in
Fig.~2;
note the steep central gradient, the change of slope at $a\sim2''$,
and the secondary minimum at $a\sim45''$ of $v(a)$.
It is of interest to remember that in the range of these observations, from
$a\simeq1/5a_e$ to $\simeq 1/2\,a_e$ along the major axis, the disk is
brighter than the bulge (Fig.~3).
With respect to preceding studies our result doubles the interval of
galactocentric distances over which $v(a)$ has been found to run constant in
NGC~3115.

\begin{figure}
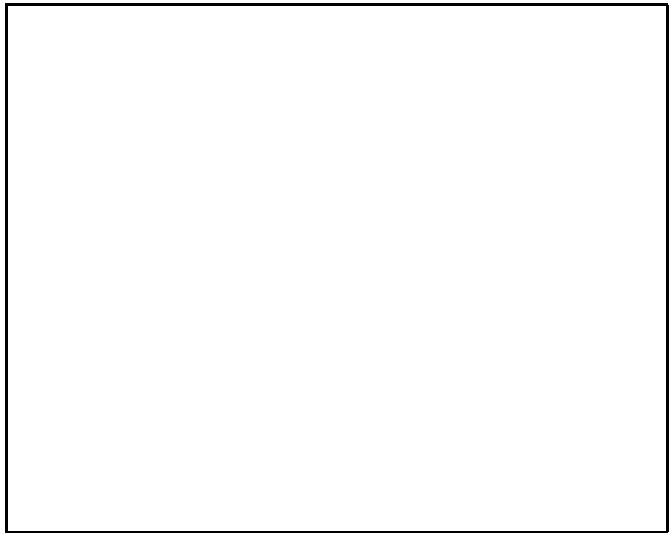

\picplace{7cm}
\caption{
Fractional surface density of light for the disk of NGC~3115 along the
direction of the major axis. Data have been taken from the photometric
study of NGC~3115 by Capaccioli, Held \eppoi\ Nieto (1987).}
\end{figure}

\subsection{The velocity dispersion curve\label{SSvel_disp}}
Mean values of the velocity dispersion $\sigma$ from the EMMI spectra, after
folding about the galaxy center and averaging with the same scheme as for the
radial velocities, are shown in Fig.~4.
The \rms\ dispersion on $\sigma$, which is very small at $a \mincir 1.4
a_e$ ($\Delta\sigma < 5$ \kms), blows up at larger galactocentric distances
(where it is also based on one side of the galaxy only).
Therefore we have discarded all data at $a>130''$;
the reason for the difference between radial velocity and velocity dispersion
curves is that the rapidly decreasing signal--to--noise (S/N) ratio affects
more severely $\sigma$ than $v$ measurements.

\begin{figure}
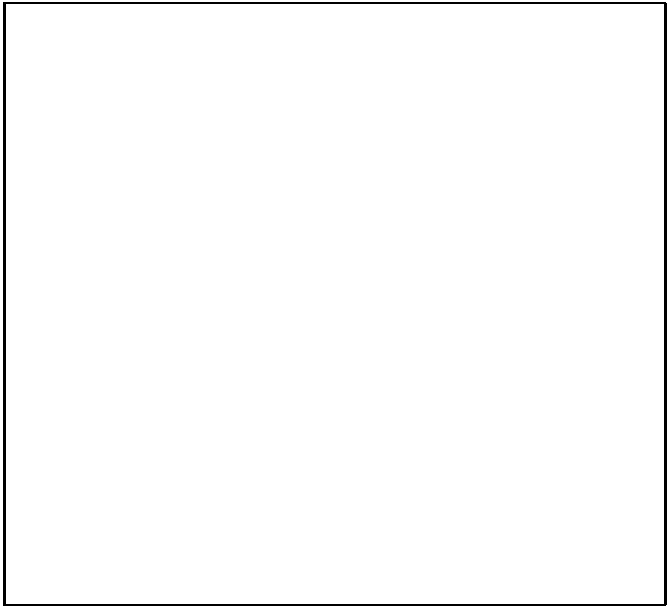

\picplace{8cm}
\caption{
Same as Fig.~1 for the velocity dispersion.}
\end{figure}

\begin{figure}
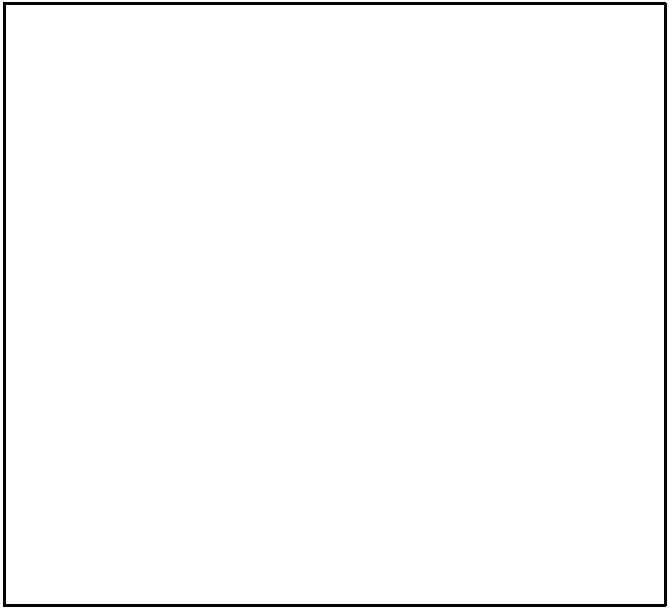

\picplace{8cm}
\caption{
Same as Fig.~2 for the velocity dispersion.}
\end{figure}

The assembly of the measurements from the various EFOSC spectra into a final
velocity dispersion curve was made easy by the sharp central cusp (whose
spatial coordinate was always found to coincide with that of the corresponding
radial velocity curve).
The individual determinations of $\sigma(a)$ were co-added and averaged using
the same scheme as adopted for $v(a)$.
The resulting values $\sigma_\ast(a)$ were still affected by this
convolution, induced by the drift of the spectra across the detector
during long exposures.
We postulate that the corrected values $\sigma(a)$ depend on the convolved
velocity dispersions through the relation\footnote{In a preliminary report
(Cappellaro \etal\ 1989, 1990) we simply assumed $\sigma=\sigma_\ast/1.5$ at
all distances, from comparison with Illingworth \eppoi\ Schechter's (1982).}
\begin{equation}
\sigma(a)=\Big[\sigma_\ast^2(a)-s_\circ^2\Big]^{1/2},
\end{equation}
where $s_\circ$ is a constant to be found by comparison with the EMMI and B\&C
velocity dispersion curves.
With $s_\circ=125\pm5$ \kms, corresponding to a $20\%$ increase in the
instrumental dispersion, we obtained an excellent agreement (Fig.~4 and 5) with
the EMMI and B\&C curves as well as with the less resolved data of Illingworth
and Schechter (1982).
Note that the matched values span the interval from $\sigma \simeq 100$ to
$300$ \kms\ and that even the central cusp of $\sigma$ is in agreement with the
high resolution data plotted in Kormendy (1987) and Kormendy \eppoi\ Richstone
(1992); see also Fig.~6.
Note also that the EMMI velocity dispersion values are in perfect agreement
with the EFOSC data over the entire range, proving that the large and somewhat
empirical correction indeed works.
An interpolated curve through the two sets of data is sampled in Table~3 and
plotted in Fig.~4.

\begin{figure}
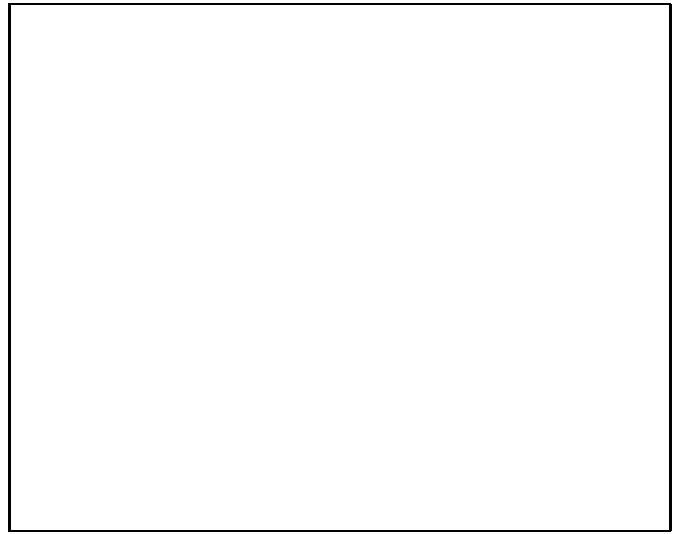

\picplace{7cm}
\caption{
Comparison of our kinematical measurements (filled squares = EFOSC,
filled circles = EMMI) with the high--resolution observations by
Kormendy \eppoi\ Richstone (1992) (open circles). The $r^{1/4}$ scale has
been chosen to facilitate comparison.}
\end{figure}

In any case, we have analysed the effects of the instrumental resolution
$\sigma_{inst}$ and of the signal--to--noise ratio for the EFOSC material.
Simulations with our reduction package show that the systematic
error on the measured value of the velocity dispersion is $\mincir 20\%$ for
$\sigma > 0.4 \sigma_{inst}$, \ie\ for $\sigma > 80$ \kms\ in our case.
We have also compared the CC package with a Fourier quotient (FQ) package
developed and kindly made available to us by Dr.\ G.\ Galletta.
In agreement with the results of Larsen \etal\ (1983), we find that CC seems
superior to FQ in measuring velocity dispersions, particularly if they are
small.
The next question is the influence of the signal--to--noise ratio;
for decreasing S/N, the measured $\sigma_{obj}$ is dominated by the noise.
Our simulations show that actually, at all $a \magcir 1.3 a_e$, the signal
of our spectra is such that we cannot reliably measure velocity dispersions
which are $\mincir 100$ \kms.
Therefore all values of $\sigma$ at $ a > 1.3 a_e$ are conservatively
discarded.

A comparison of our measurements with the same literature data of
Fig.~2 is provided in Fig.~5. Furthermore, comparison with
Kormendy \eppoi\ Richstone (1992) is provided in Fig.~6.

\section{Theoretical models}
The analysis of the data presented above requires fitting the light profiles
and the rotation velocity curve with analytical forms.
The photometry (\CHN\ 1987) was fitted to the projection of a Jaffe law which,
for ease of computation of the potential gradients, was modified as follows:
\BGE\label{jaffe}
\rho(m^2) = \rho_0 \frac{R_J^4}{m^2(m^2 + R_J^2)}
\EDE
where
\BGE
m^2 \equiv R^2 + \frac{z^2}{c^2}
\EDE
with $c$ the intrinsic flattening of the bulge, and $R_J$ a parameter to be
fitted.
This form of the density distribution has the same asymptotic behavior as
Jaffe's law (Jaffe 1983) for small and large $m$'s, but leads to simpler
potential gradients.
Figure~7 shows the fit to the observed minor--axis light profile, which
represents the pure bulge since NGC~3115 is close to edge--on view.
The reduction to the major axis, taking into account the variable flattening
of the bulge isophotes (Table~8 in \CHN\ 1987), is also given;
the excess light in the observed profile is entirely due to the edge--on disk
(\MC, Vietri \eppoi\ Held 1988).
The best fitting Jaffe radius for the major axis is $R_J = 92''$, almost
coincident with the effective radius of the bulge (Table~1).

\begin{figure}
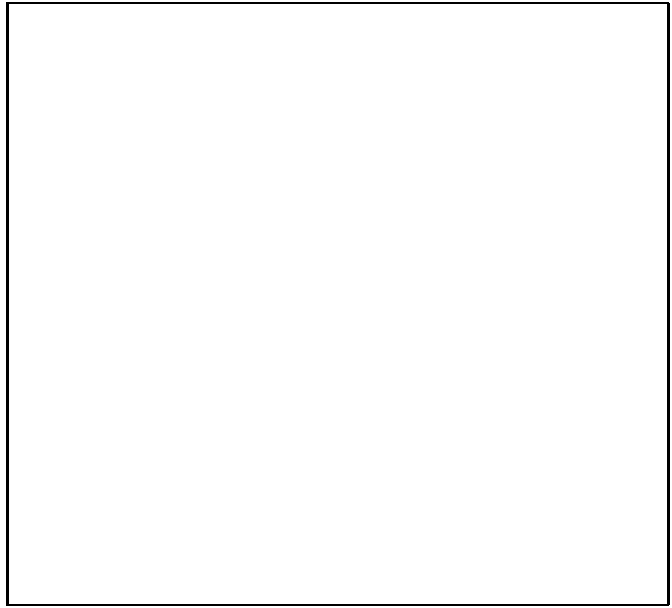

\picplace{8cm}
\caption{Luminosity and flattening profiles and relative growth curve for
NGC~3115 (from \CHN\ 1987). The open circles reproduce the observed light
profiles along the major and the minor axis; neither of them has been
corrected for Galactic absorption ($A_B=0.1$, after Burstein \eppoi\ Heiles
1982). The solid lines are the Jaffe--model fit to the minor axis and its
reduction to the major axis ($R_J = 92''$), done using the ellipticity profile
of the bulge. The excess of the observed major axis light profile over the
model is due to the presence of the exponential disk (\MC, Vietri \eppoi\ Held
1988). The dotted line is the radial run of the apparent ellipticity of the
bulge; the dashed line gives the relative growth of the luminosity,
$L/L_T$.}
\end{figure}

An analytical approximation to the observed rotation velocity curve,
\BGE\label{rotfit}
v_{rot} = v_\infty\frac{R}{R + R_c}
\EDE
is shown in Fig.~8, with $v_\infty=260$ \kms, and $R_c = 2\secondip5$.

The question we are interested in is whether NGC~3115
has dark matter within the radius for which
observational data are available.
Thus, in a first approximation, we build self-gravitating, constant \ML\
models,
which must satisfy the constraints provided by the photometric and
spectroscopic observations.

Models of ellipsoidal systems based on Jeans' equations (Binney \eppoi\
Tremaine 1987) can be built provided that extra assumptions are introduced to
close the system.
One reasonable such assumption, discussed by Binney, Davies \eppoi\ Illingworth
(1990; hereafter BDI), is that the distribution function depends only on the
two classical integrals of motion in an axially symmetric potential, the
energy $E$ and the $z$--component of the angular momentum $L_z$.
Under such circumstances, the velocity dispersion along the $z$--axis equals
that along the $R$--axis, thus closing the system of Jeans' equations.
A variation of the method of BDI is discussed in the Appendix.
It shall be borne in mind that all of our models depend also on the accuracy
of the assumption that the galaxy is oblate.
It is well--known that NGC~3115 has a nearly perfectly edge--on disk
(\MC\ \etal\ 1988);
this, and an argument based on its resemblance to NGC~3379 (\MC\ \etal\
1991), lead us to expect that such an assumption should be roughly correct.
In this case, the apparent flattening of NGC~3115 coincides with its intrinsic
flattening, and can immediately be inserted into Eq.~\ref{jaffe} and in the
computation of the potential gradients.

The effect of the disk is included: it has been modelled as
an exponential disk with zero thickness, scale--length $r_\circ = 25
\secondip .5$ and \ML\ = 7 (Capaccioli, Vietri \eppoi\ Held 1988).
Comparison with models without disk shows that its dynamical influence
is negligible at all radii, consistent with its accounting for only
$\simeq 6\%$ of the total light. For galaxies with massive haloes,
the presence of a disk can only make matters worse: in fact, the presence of
the disk reduces the bulge contribution in the central parts, and thus
also reduces the predicted rotation and velocity dispersion curves in the
outer parts, where the disk contribution has died away.

The main result of this analysis is shown in Fig.~9.
Here, the observed velocity dispersion along the line of sight is compared to
the prediction of the model.
Two different curves are presented;
the dashed one includes a central mass concentration (black hole as in
Kormendy 1987, where the relevant parameters are given), with which the fit to
the kinematical data in the innermost regions of the galaxy is improved.
The presence of this black hole, whose dynamical importance in the outermost
regions is of course negligible, still indirectly affects our modeling, since
it reduces the amount of matter needed to fit the observational data, and thus
reduces the \ML\ of the visible matter, assumed independent of radius.
In fact we see that the model without the black hole fits the outer data
somewhat better than the model with the black hole.
We obtain $\ML=6.5$ in solar units for the model without black hole, and
$\ML=6$ for the model with black hole.

However, the most interesting feature of Fig.~9 is that none of the models
can follow the roughly flat behavior of the observed velocity dispersion for
$R\magcir0.5\,R_J$.
It is easy to understand that, given the above assumptions, our data must lead
to a physically unacceptable model.
In fact, we have no evidence for a decrease in either the rotation velocity or
velocity dispersion, while in an isotropic self--gravitating Jaffe model both
{\it must\/} decrease over the radial range observed by us.
The rotational velocity ($\equiv(-R\partial\phi/\partial R)^{1/2}$), which
of course is an upper limit to the observed one because of the contribution of
the velocity dispersion to the model's support against self--gravity, is shown
in Fig.~8 (dashed line) for the assumed modified Jaffe model with a run of
ellipticity as in Table~8 of \CHN\ (1987).
It can be easily seen that such rotational velocity decreases, contrary to our
evidence.
Thus our data alone are sufficient to exclude a self--gravitating
model with constant \ML.

\begin{figure}
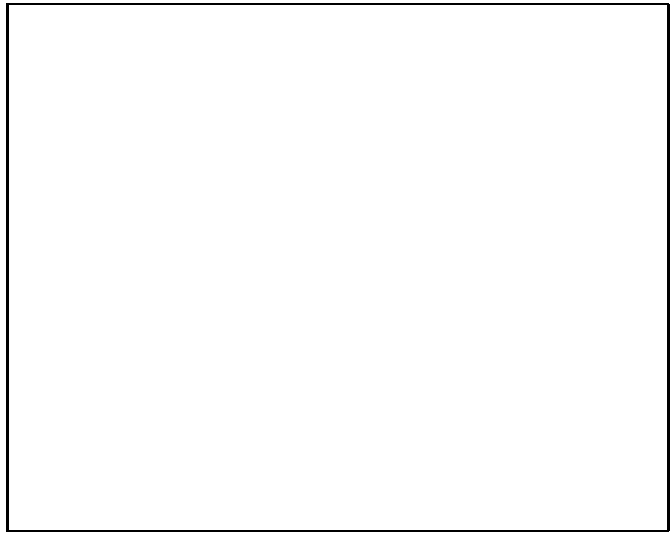

\picplace{7cm}
\caption{
Rotation curve for NGC~3115.
The data points are those of Fig.~1 and ~2.
The solid line plots the approximation of Eq.~4, with $R_c=2\secondip5$.
The dashed line is the projected rotation velocity for a Jaffe model with zero
velocity dispersion and with the run of ellipticity variations taken from
Fig.~13 of \CHN\ (1987).}
\end{figure}

\begin{figure}
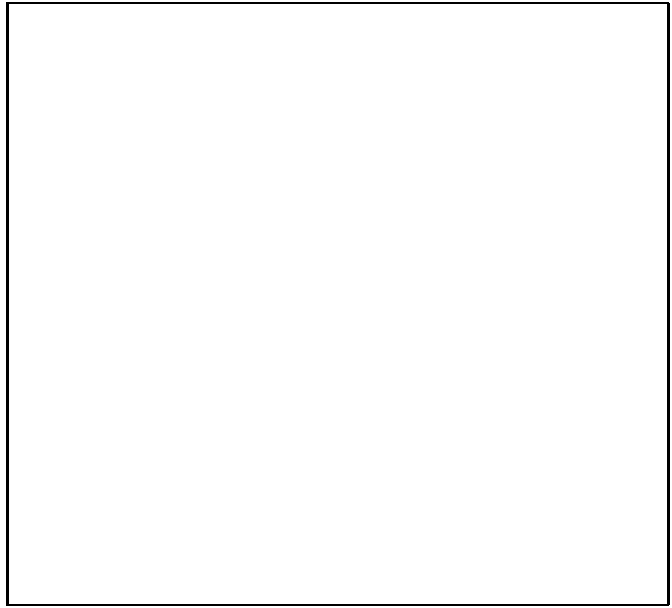

\picplace{8cm}
\caption{
Velocity dispersion along the light of sight measured along the major axis
Data are from Fig.~4 and ~5.
The solid line is a fit with $\ML= 6.5$ in solar units, and no central black
hole;
the dashed line is another fit with $\ML = 6$ and a central black hole with
parameters taken from Kormendy (1987).
Velocity units are \kms; radii are in units of the Jaffe radius $R_J = 92''$.}
\end{figure}

On the basis of our models it is less easy to estimate the magnitude of the
mass discrepancy in the outer parts.
Fig.~9 shows that both curves predict $\sigma_V^2 = 0$ for observable radii,
$R\simeq 0.8 R_J$ for both models.
Clearly, such results depend on our assumptions, in particular on having
assumed the galaxy to be oblate and $\sigma_z^2 = \sigma_R^2$.
It may be possible, by relaxing such constraints, to build a better--fitting
model.
Still we would like to point out that, from our data, there is a minimum \ML\
for any model of this galaxy.
In fact, we know that, for $R=180''$, it is $v_{rot} \approx 260$ \kms.
A {\it minimum\/} \ML\ model will have from there on $\sigma_z = \sigma_{\phi}
=\sigma_R = \sigma_V = 0$; then the minimum \ML\ can be estimated by equating,
at that galactocentric distance, centrifugal and gravitational accelerations.
We obtain $(\ML)_{min} = 10$ in solar units.
It is well--known that the inner rotation curve ($R<0.5 R_J$) can be fitted
with $\ML=6$ (Illingworth \eppoi\ Schechter 1982, Rubin \etal\ 1980),
so that we now find a discrepancy of {\it
at least\/} a factor of 2 in \ML\ while moving from $R\simeq0.5R_J$ to
$R\simeq2R_J$. It is easy to understand this from Fig.~7, which shows
that $L$ changes by only $20\%$ from $R=90''$ to $R=180''$,
while the minimum mass $M_{min}(R) \propto R\,v_{rot}^2$ doubles over the
same range, for our flat rotation curve (Fig.~8).

\section{Summary and conclusions}
This is the first paper of a series based on deep major--axis spectra of
bulge--dominated edge--on S0's, obtained by using EFOSC in combination with the
ESO 3.6\,m telescope and, in this one case only, EMMI at NTT as a check for
our unconventional use of EFOSC.
Aim of this series is to report on direct kinematical evidence of dark matter
haloes within the optical images of galaxies related to ``disky'' E's, \ie\
that half of all normal ellipticals which do not offer any other way for
probing the occurrence of dark haloes.
We summarize briefly our conclusions:

1.\ \ We have measured the rotation curve
of NGC~3115 out to a galactocentric distance $\simeq2a_e$, where the surface
brightness is $\sim 20\%$ of that of the night sky.
It increases steeply in the galaxy core,
then remains constant (within $10\%$) at $\vass = 260$ \kms\ out to $2a_e$.
Similarly, the observed velocity dispersion peaks sharply at the center (with
a maximum of 325 \kms), decreasing outwards till it flattens out at
the level of $\sim 100$ \kms;
reliable measurements are limited to $a < 1.3a_e$.
There is a marginal indication from the EMMI spectra that $\sigma$ decreases
for $a \magcir 2'$.

2.\ \ Our simple--minded models, assuming
oblateness, constant \ML, and $\sigma_z^2 = \sigma_R^2$, predict {\it
vanishing\/} velocity dispersions along the line of sight, well within our
observed range.
We have argued in a qualitative way that fitting our data with \ML=const,
is unlikely, even when the afore--mentioned assumptions are relaxed.
Also, we have shown that a simple argument implies that, within our observed
range, \ML\ increases by {\it at least\/} a factor of $\sim2$, from $\ML=6$ in
the center to $\ML=10$ at our outermost radius ($1.8 a_e$).

3.\ \ There is a simple argument to show that the disk is not only unimportant
from the photometric point of view, but also from the dynamical point of view.

\bigskip\noindent
{\it Acknowledgements.}\ \ We wish to thank Drs. Ralf Bender and Giuseppe
Galletta for letting us use their software, and Professor
James Lequeux for a careful reading of the manuscript, and for pointing
out an error in a previous version.

\bigskip
\section*{Appendix}
We summarize here the method used to analyze the kinematical data, which is
similar to that developed by BDI.
We model the galaxy as a modified Jaffe model (MJM, from now on), with density
constant on self--similar ellipsoids
\BGE
m^2 \equiv x^2 + y^2 + \frac{z^2}{c^2}
\EDE
($c$ is the true axial ratio) and density profile
\BGE
\label{dens}
\rho = \frac{1}{2\pi^2 c\,(R_2 + R_1)(m^2 + R_2^2)(m^2 + R_1^2)}
\EDE
where the total mass $M = 1$ and $R_2 \gg R_1$ are two constants.
The introduction of a second scale parameter ($R_1$, much smaller than any
observed radius) is used to simplify the computation of the gravitational
potential.

The model that we are trying to fit to the observational data must satisfy
Jeans' equations.
Now we assume that the system that emits the light is oblate and
self--gravitating (\ie\ it has constant mass-to-light ratio), and has velocity
dispersion $v_z^2$ along the $z$--axis equal to that along the $R$--axis,
$v_R^2$.
The first two assumptions are obvious, the third one is discussed by BDI.
Then Jeans' equations for this model in cylindrical coordinates are
\BGE
\frac{\partial(\rho\sigma_R^2)}{\partial R}+\rho\left(\frac{\sigma_R^2-
v_{\phi}^2}{R}+\frac{\partial(\phi+\phi_{ext})}{\partial R}\right)=0
\EDE
\BGE
\frac{\partial(\rho\sigma_z^2)}{\partial z}+\rho\frac{\partial(\phi+
\phi_{ext})}{\partial
z}=0
\EDE
Here $\phi$ is the gravitational potential self--consistently engendered by
the light--emitting system, and is given for reference below in explicit
form in the case in which $c$ is independent of radius, and $\phi_{ext}$
is the disk potential, computed for an infinitely thin exponential
disk (Binney \eppoi\ Tremaine 1987), with the parameters described
in Section 4.
Clearly, since $\rho$ and $\phi$ are supposed known, these two coupled
equations become a system of two unknowns, $\sigma_z^2 \equiv \sigma_R^2$ and
$v_{\phi}^2$, which can be readily solved numerically.

The total velocity dispersion $v_{\phi}^2$ thus determined includes
contributions from both the rotational velocity and the true velocity
dispersion:
\BGE
v_{\phi}^2 = v_{rot}^2 + \sigma_{\phi}^2
\EDE
On the other hand, observations give us $v_{rot}^2$, which can be subtracted
from $v_{\phi}^2$ to yield $\sigma_{\phi}^2$.
We now have all the principal components of the velocity dispersion tensor,
and
can thus project them (see for instance Binney \eppoi\ Mamon 1982) to give
$\sigma_V^2$, the velocity dispersion along the galaxy major axis, and this
can
be compared with the observed quantity $\sigma_{obs}^2$.
However, we still have the extra freedom of an unknown multiplicative
constant, essentially the \ML\ of the galaxy.
In keeping with the spirit of the model, which is self-gravitating and thus
ought to reproduce the kinematics of the {\it inner} regions of the galaxy, we
fix the \ML\ by demanding that the theoretically determined
$\sigma_V^2$ fits $\sigma_{obs}^2$ in the inner regions of the galaxy.

Now our model has no free parameters left, and ought to reproduce
$\sigma_{obs}^2$ in the whole observable region.

The main difference between our method and that of BDI lies in the fact that
we treat the surface photometry as if the light profile where of the Jaffe
type everywhere, while they use a Lucy algorithm to deproject the
two--dimensional image (or close to this: they have light-emission profiles
along a number of PAs).
We compensate for this by taking for the axial ratio $c$ not a constant, but
the actual observed run with $R$.
Since the determination of $c$ in standard packages essentially requires
observations along 3 PA's (we discard the term \mbox{$\propto \cos4\theta$}),
this compares reasonably well with BDI, who had data along 7 PA's.
Also, since NGC~3115 is seen edge--on, the axial ratio $c$ is not a free
parameter for us, since $c = c_{obs}$.
Lastly, BDI do not subtract the observed velocity rotation curve like we do.
But since in this case the data for $\sigma_{obs}^2$ and $v_{rot}$ are
coextensive, the method adopted here seemed the most appropriate one.

We give here for reference the gradients of the gravitational potential
engendered by the flattened MJM, for the case of constant axial ratio $c$.
In these formulae, some simplification is possible through the fact that the
density (Eq. \ref{dens}) can be rewritten as
\BGE
\rho = \frac{1}{2\pi^2 c\ (R_2 + R_1)} \left( \frac{1}{m^2 + R_1^2} -
\frac{1}{m^2 + R_2^2} \right)
\EDE
Such potential is available through well--known implicit formulae (\eg\ Binney
\eppoi\  Tremaine 1987).
We define $R^2 \equiv x^2 + y^2$, take $G = 1$, and omit the algebra, to find
\BGE
\frac{\partial \phi}{\partial R} = - \frac{2R}{\pi\ (R_2 - R_1)} \Big[ X(R_1) -
X(R_2) \Big]
\EDE
where
\BGE
X(R_i) \equiv \frac{1}{R_i^2} \Big[ g(\tau_1,\tau_2,1) + g(\tau_2,1,\tau_1) +
g(1,\tau_1,\tau_2) \Big]
\EDE
Here we have defined $\tau_1 < \tau_2$ as the {\it moduli} of the solutions of
\BGE
R_i^2 \tau^2 + \Big[(1+c^2)R_i^2 + z^2 + R^2\Big]\tau +
R_i^2c^2 + z^2 + R^2c^2 = 0
\EDE
(which, of course, depend on $R_i$), and
\BGE
g(\tau_1, \tau_2, x) \equiv \frac{(1- c^2)^{1/2}}{(\tau_2 - \tau_1)
(x-\tau_1)} \left[ \frac{\pi}{2} - \arctan \frac{c}{(\tau_1-c^2)^{1/2}}
\right]
\EDE
Analogously, we find
\BGE
\frac{\partial \phi}{\partial z} = - \frac{2z}{\pi (R_2 - R_1)} \Big[ Y(R_1) -
Y(R_2)\Big],
\EDE
with
\BGE
Y(R_i) \equiv \frac{1}{R_i^2(\tau_1 - \tau_2)} \Big[ h(\tau_2) - h(\tau_1)
\Big],
\EDE
\BGE
h(\tau) \equiv \frac{\frac{\displaystyle \pi}{\displaystyle 2} -
\arctan \frac{\displaystyle c}{\displaystyle(\tau - c^2)^{1/2}}
}{{(\tau - c^2)^{1/2}}}
\EDE
and the $\tau$'s are defined as above.

\end{document}